\documentclass[preprint2,twoside]{hwo}
\usepackage{enumitem}
\usepackage[table,dvipsnames]{xcolor}

\usepackage{graphicx}
\usepackage[table]{xcolor}

\input{hwo.h}

\begin{document}

\title{\textbf{\LARGE Assessing Ocean World Habitability with HWO}}
\author {\textbf{\large R.J. Cartwright$^1$, L. Quick$^2$, M. Neveu$^{2,3}$, T.M. Becker$^4$, U. Raut$^4$, \\ J.C. Castillo-Rogez$^5$, K.L. Craft$^1$, G.L. Villanueva$^2$}
\vspace{0.3 cm}\affil{$^1$\small\it Johns Hopkins University Applied Physics Laboratory, Laurel, Maryland, USA}
\vspace{0.1 cm}\affil{$^2$\small\it NASA Goddard Space Flight Center, Greenbelt, Maryland, USA}
\affil{$^3$\small\it University of Maryland, Baltimore, Maryland, USA}}
\affil{$^4$\small\it Southwest Research Institute, San Antonio, TX, USA}
\affil{$^5$\small\it Jet Propulsion Laboratory California Institute of Technology, Pasadena, California, USA}

\author{\footnotesize{\bf Endorsed by:}
Chloe Beddingfield (Johns Hopkins University, Applied Physics Laboratory), Katherine Bennett (Johns Hopkins University), Kara Brugman (Arizona State University), Ligia Coelho (Cornell University), Leigh Fletcher (University of Leicester), Luca Fossati (Institut für Weltraumforschung, ÖAW), Christopher Glein (Southwest Research Institute), William Grundy (Lowell Observatory), Heidi Hammel (Association of Universities for Research in Astronomy), Kevin Hand (Jet Propulsion Laboratory, California Institute of Technology), Matthew Hedman (University of Idaho), Natalie Hinkel (Louisiana State University), Bryan Holler (Space Telescope Science Institute), Chris Impey (University of Arizona), Adam Langeveld (Johns Hopkins University), Eunjeong Lee (EisKosmos, CROASAEN), Shannon MacKenzie (Johns Hopkins University, Applied Physics Laboratory), Stefanie Milam (NASA Goddard Space Flight Center), Faraz Nasir Saleem, John Rosenfield, Lorenz Roth (KTH Royal Institute of Technology), Pablo Santos-Sanz (Instituto de Astrofísica de Andalucía, CSIC), Melinda Soares-Furtado (University of Wisconsin), Federico Tosi (Istituto Nazionale di Astrofisica, INAF), Elizabeth Turtle (Johns Hopkins University, Applied Physics Laboratory) 
}



\begin{abstract}
The instrument payload of the future Habitable Worlds Observatory (HWO) will span a wide range of wavelengths, including the ultraviolet (UV) region that cannot be easily accessed from the ground ($<$ 350 nm). Along with its primary mission to characterize the habitability of candidate exo-Earths, HWO will be well suited for observations of potentially habitable icy ocean worlds in our Solar System, in particular with an integral field spectrograph (IFS). Here, we discuss future HWO observations of ocean worlds including Ceres, Europa, Enceladus, Ariel, and Triton. We explore the observational requirements for capturing ongoing and sporadic geyser activity and for measuring the spectral signatures of astrobiologically-relevant compounds, including water, salts, organics, and other bioessential components. We consider the key observing requirements for an IFS, including wavelength coverage, resolving power (R), angular resolution, and field-of-view (FOV). We also outline some of the potential measurements that would define incremental, substantial, and breakthrough progression for characterizing habitability at ocean worlds, primarily focusing on UV and visible (VIS) wavelengths (90 -- 700 nm). Our investigation concludes that a UV/VIS IFS on HWO could make some groundbreaking discoveries, in particular for detection and long-term monitoring of geyser activity and interior-surface exchange of components critical for understanding habitability at ocean worlds. 
\vspace{0.6 cm}
\end{abstract}

\section{Introduction}
\vspace{-0.1 cm}
\textit{Driving science question: Do large icy bodies and moons harbor habitable environments in their interiors?} Many dwarf planets and icy moons in the Solar System could have discrete pockets of brines within their ice shells, and/or global saline oceans wedged between their rocky mantles and icy exteriors \citep[e.g.,][]{hussmann2006OWs,hussmann2010OWs, nimmo2016OWs, lunine2017OWs, hendrix2019ROW}. Liquid water in the interiors of these ``ocean worlds'' could reach their surfaces via geologic conduits and/or other geophysical processes, such as is likely the case on Jupiter’s moon Europa \citep[e.g.,][]{schmidt2011Europachaos, hesse2022Europaexchange, chivers2023Europaexchange, daubar2024Clippergeol} and the main belt dwarf planet Ceres \citep[e.g.,][]{nathues2020Cerescryo, raymond2020Ceresimpact}, depositing salts, organics, and other compounds that can be used to decipher internal habitability. At Saturn’s moon Enceladus, ocean material is erupted into space by multiple geysers \citep[e.g.,][]{porco2006EnceladusISS, porco2017Enceladusmicrobes, waite2006EnceladusINMS, waite2009EnceladusNH3, waite2017EnceladusH2, hedman2009EnceladusVIMS, hansen2011EnceladusUVIS, goldstein2018Enceladusplume}. Studying the chemical composition of geyser material provides invaluable insight into ocean environmental properties and habitability. It is important to confirm which icy bodies could harbor habitable environments based on the presence of liquid water, internal energy sources, and bioessential components (i.e., ``CHNOPS'' bearing species), as well as assessing ongoing endogenic activity that amplifies fluxes of such energy sources and materials \citep[e.g.,][]{cockell2016habitability,hand2020OWhabitability}.

\textit{Connections to Exoplanet Habitability.} Habitable environments in our Solar System that are amenable to characterization by remote sensing extend well beyond the Goldilocks Zone and past the H$_2$O snowline. By extension, the habitable zones (HZs) of exoplanets with surface oceans may extend beyond the H$_2$O snowline of their host stars. These ``cold ocean'' exoplanets may have partly or fully frozen surfaces enveloped by atmospheres, such as the likely super-Earth LHS 1140b \citep{cadieux2024LHS1140bJWST}. Such cold ocean exoplanets could retain habitable environments beneath their icy exteriors, perhaps analogous to Europa and other icy ocean worlds in our Solar System \citep[e.g.,][]{kane2021exoplaneticy,quick2020exoplanetcryo,quick2023exoplanetcryo}. Similarly, icy exomoons may have habitable saline oceans beneath their icy exteriors, sustained by radiogenic and tidal heating sources \citep{tjoa2020exomoonhab}. 

High quality spectral measurements of ocean worlds in our Solar System, collected over multi-decadal timelines, could prove to be an important source of knowledge for understanding the key tracers of habitability on cold ocean exoplanets and icy exomoons. In particular, measuring the spectral properties of ocean worlds at wavelengths corresponding to those covered by HWO’s future coronagraph(s) (\(\sim\)250 -- 1700 nm) are key for developing potentially useful analogs for icy exoplanets and exomoons.

\subsection{Summary of Observing Requirements for Ocean Worlds}
\begin{itemize}
  \item{\textit{Wavelength coverage:} Ideally EUV-NIR, 50 -- 5000 nm; minimum FUV-VIS, 90 -- 700 nm.}
  \item{\textit{Resolving Power (R):} Ideally $\sim$ 10,000; minimum of a few thousand, spanning UV-NIR.}
  \item{\textit{Signal-to-Noise Ratio (SNR):} $>$ 100 across UV-NIR.}
  \item{\textit{Angular resolution:} Ideally $\le$ 0.015''/pixel, minimum $\sim$0.02''/pixel over UV/VIS wavelengths.}
  \item{\textit{Field of view (FOV):} Ideally 10''x10'', minimum 3''x3'', with a high dynamic range and saturation mitigation procedures for observations of bright objects (e.g., Ceres and Europa). } 
  \item{\textit{Observation scheduling:} UV/VIS and NIR spectroscopy must be contemporaneous for ocean world science, sampling the full wavelength range within the same $\sim$3-hour window. Although not required, simultaneous observations across this wavelength range could have serendipitous benefits for characterizing transient phenomena (such as Europa geysers). UV/VIS wavelength coverage is most critical, with NIR wavelengths perhaps provided by a later generation IFS, added to HWO during a post-launch servicing run. Furthermore, follow up observations of the same ocean world target should be conducted over multi-decadal timescales (hereafter referred to as ``long-term monitoring'').}
\end{itemize}

\textit{IFS or MOS?} As demonstrated by decades of spacecraft data collected at Ceres, Jupiter, Saturn, and Pluto, as well as recent observations of the Galilean moons conducted with the James Webb Space Telescope (JWST), imaging spectroscopy is essential for characterizing ocean worlds and other targets of Solar System science cases \citep[e.g.,][]{carlson1996GmoonsNIMS, brown2006EnceladusVIMS, desanctis2015CeresNH4, grundy2016PlutoLEISA, villanueva2023EuropaJWST, villanueva2023EnceladusJWST, cartwright2024CallistoJWST, cartwright2025EuropaJWST}. To this end, many Solar System science cases require IFS measurements to achieve their objectives. An IFS would be ideal for measuring the spectral properties of extended objects like Jupiter and its large moons, and other spatially resolved objects like Ceres and Titan. Furthermore, an IFS would be well suited for measuring geysers and exospheres, which require spectroscopic characterization over a range of radial distances from an ocean world's surface (see E-ring and Enceladus geysers in Figure~\ref{Enceladus}). 

In contrast, a Multi-Object Spectrograph (MOS), with numerous, typically small slitlets (FOVs $<$ 0.5''x0.5''), is not ideal for mapping the distribution of salts and other components across the resolved disks of Europa and the other Galilean moons (angular diameters $\sim$0.9 -- 1.7''). Measuring the radial extent of extended geyser eruptions and transient exospheres would also be difficult with MOS slitlets. Keeping non-sidereal, Solar System targets near the center of small MOS slitlets is more challenging and requires additional set up and observation overheads compared to the larger FOV of an IFS. For example, JWST's NIRSpec, operating in MOS mode, only provides the option for a pseudo-long slit for use with extended moving targets, but the utility of this setup is questionable due to potential mechanical issues (stuck open/closed shutters), the loss of spatial information, and possible flux leakage through the micro-shutter arrays due to the placement of large and bright targets directly on the array. Thus, no such MOS mode observations have been approved for Solar System targets through the first 4 JWST cycles, and NIRSpec's integral field unit (FOV 3''x3'') is the preferred option.  

\subsubsection{IFS Rationale Summary.} An IFS that spans the UV and VIS (90 -- 700 nm), with an R $\sim$ 3000 and capable of angular resolutions that improve upon the Hubble Space Telescope (HST) and JWST ($\le$ 0.02''/pixel), is critical for understanding habitability at ocean worlds across the Solar System. Wavelength regions observed simultaneously will provide key data on temporal changes in e.g., auroral activity at the Galilean moons. Furthermore, long-term monitoring would provide crucial information on seasonal and diurnal variations in geyser and exospheric activity, volatile migration, and changes in charged particle fluxes. NIR wavelength coverage, perhaps provided by a later generation IFS operating at ambient temperatures, would provide highly complementary data on ocean world activity, such as long-term monitoring of H$_2$O and CO$_2$ gases erupted by geysers and residing in transient exospheres, which are obscured by telluric contamination from Earth's atmosphere.

\section{Science Objectives}
The primary science objectives are to determine the surface and atmospheric composition of confirmed and candidate ocean worlds to assess ongoing geologic activity and potential interior-surface material exchange, the presence of compounds indicative of formation in liquid water, chemical sources of energy, and major bioessential elements (CHNOPS). Physical parameters and desired HWO observations required to achieve our science objectives are summarized in Tables ~\ref{Table 1} and ~\ref{Table 2} in section 4.

\textit{Why the obsession with salts?} At rock-water interfaces, H$_2$O molecules leech ions away from silicates (Na, Ca, K, Cl, SO$_4$, etc.) to form saltwater. In environments where water is removed due to freezing and/or vaporization (e.g., by depressurization during ascent to the surface), saltwater forms brines that eventually precipitate salts \citep[e.g.,][]{marion2005effects, castillo2018Ceresevol, buffo2020Europaocean, wolfenbarger2022OWice, naseem2023OWsalts}. Ocean worlds with internal liquid water layers might form salts at the boundary between their oceans and rocky mantles or in porous cores \citep[e.g.,][]{zolotov2007OWsalts, zolotov2001Europasalts, glass2022OWhabitability}. Common salts like NaCl and CaCO$_3$ usually form in aqueous environments. The presence of these and other salts on the surfaces of ocean worlds is therefore a potential tracer of slow processing of silicate material in a subsurface ocean \citep{castillo2022OWcarbonates}. 

Internally-derived salts must then be delivered to the surfaces of ocean worlds by ongoing or geologically-recent endogenic activity, such as tectonism and cryovolcanism \citep[e.g.,][]{quick2022Europacryo, lesage2025Europacryo}. Alternatively, geologically-older salts that are trapped in ice shells might be exposed by impact cratering and mass wasting processes. Once delivered to the surfaces of icy bodies, studying salt composition and concentration can be used to decipher ocean conditions such as acidity and salinity \citep{glein2015EnceladuspH, sekine2015Enceladusocean, castillo2018Ceresevol, glein2025phosphates}, providing first-order constraints on habitability.

\subsection{Measurements (UV, VIS, NIR spectroscopy, \\50–5000 nm)}  

\textit{Surfaces.} Characterize the spectral signature of salts (NH$_4$, CO$_3$, SO$_4$-bearing, etc.), trapped volatiles and ices (H$_2$O, H$_2$O$_2$, NH$_3$, O$_2$, O$_3$, CO$_2$, CO, etc.), simple organics (e.g., CH$_4$), and long-chain organic residues with numerous CH, CO, and CN bonds.

Irradiated carbonaceous residues that likely persist in dark material on ocean world surfaces can exhibit broad absorption features and peaks, deep in the EUV (between $\sim$50 and 150 nm; \citealt{hendrix2016carbonUV, applin2018carbonUV}). H$_2$O exhibits a diagnostic absorption feature in the FUV, between 150 and 200 nm \citep{warren2008H2OUV,hendrix2010EnceladusUV, hendrix2018SmoonsUV, hendrix2019lunarH2O}, which can be leveraged to understand radiation processing, contaminant mixing, porosity, and other surface regolith properties.

\textit{Geysers and exospheres.} Measure emission and absorption features for molecular gases in the UV, VIS, and NIR, including H$_2$O, O$_2$, CH$_4$, CO$_2$, and CO. In particular, measuring H$_2$O vapor lines between 2600 and 2700 nm and CO$_2$ lines between 4200 and 4300 nm is crucial for understanding potential outgassing at ocean worlds \citep[e.g.,][]{villanueva2023EuropaJWST, villanueva2023EnceladusJWST, villanueva2023JWSTera, cartwright2024CallistoJWST, bockelee2024GanymedeCO2gas}. These gases are not easily measured from the ground due to strong telluric bands that effectively make Earth’s atmosphere opaque in these wavelength regions.

Measure emission lines for ions derived from salts and other ocean-derived minerals, including: H, C, O, Na, Mg, S, Cl, K, and Ca. Many of these ions express strong lines in the UV and VIS, with H exhibiting a series of lines deep in the FUV, between 90 and 125 nm. 

\subsection{Why HWO?}
The Extremely Large Telescopes, coming online in the 2030’s, will be able to collect unprecedentedly high spatial resolution and SNR data for icy ocean worlds like Europa \citep{wong2020ELT}, as well as high SNR data of cold ocean exoplanets. However, strong absorption by O$_3$, H$_2$O, CH$_4$, CO$_2$ and other gases in Earth’s atmosphere effectively prevents spectroscopic observations at wavelengths $<$ 350 nm, between 2550 and 2850 nm, and between 4180 and 4500 nm. Thus, studying ion emission lines in the UV and H$_2$O vapor, CO$_2$ gas, and various solid-state ices in the NIR requires space-based facilities. 

Although JWST is currently studying these gases and volatile ices on icy ocean worlds using NIR spectroscopy, this facility will no longer be operational once HWO has seen first light in the early 2040s. Monitoring of atmospheric changes, geyser activity, and volatile migration will not be possible without access to NIR wavelengths on a space telescope.

HWO, as a serviceable facility,  will be ideally positioned to conduct long-term monitoring of astrobiologically-relevant compounds over seasonal and longer timescales and conduct rapid response observations of transient phenomena like geysers. Furthermore, HWO's planned UV wavelength coverage, sensitivity, and angular resolution will represent impressive improvements over the Hubble Space Telescope (HST), the current premier UV-capable platform. Thus, HWO will be well positioned to study the habitability of exoplanets and icy bodies across the Solar System. 

\section{Specific Targets of Interest}
To help support the case for studying habitability of icy body interiors, we discuss some of the most promising confirmed and candidate ocean worlds that exhibit evidence for ocean-derived surface deposits, geyser activity, volatile-rich exospheres, and/or geologic evidence for interior-surface material exchange. All physical parameters that ideally would be measured by HWO are summarized in Table 1.

\subsection{Ceres}

        \begin{figure}[b!]
			\includegraphics[width=\columnwidth]{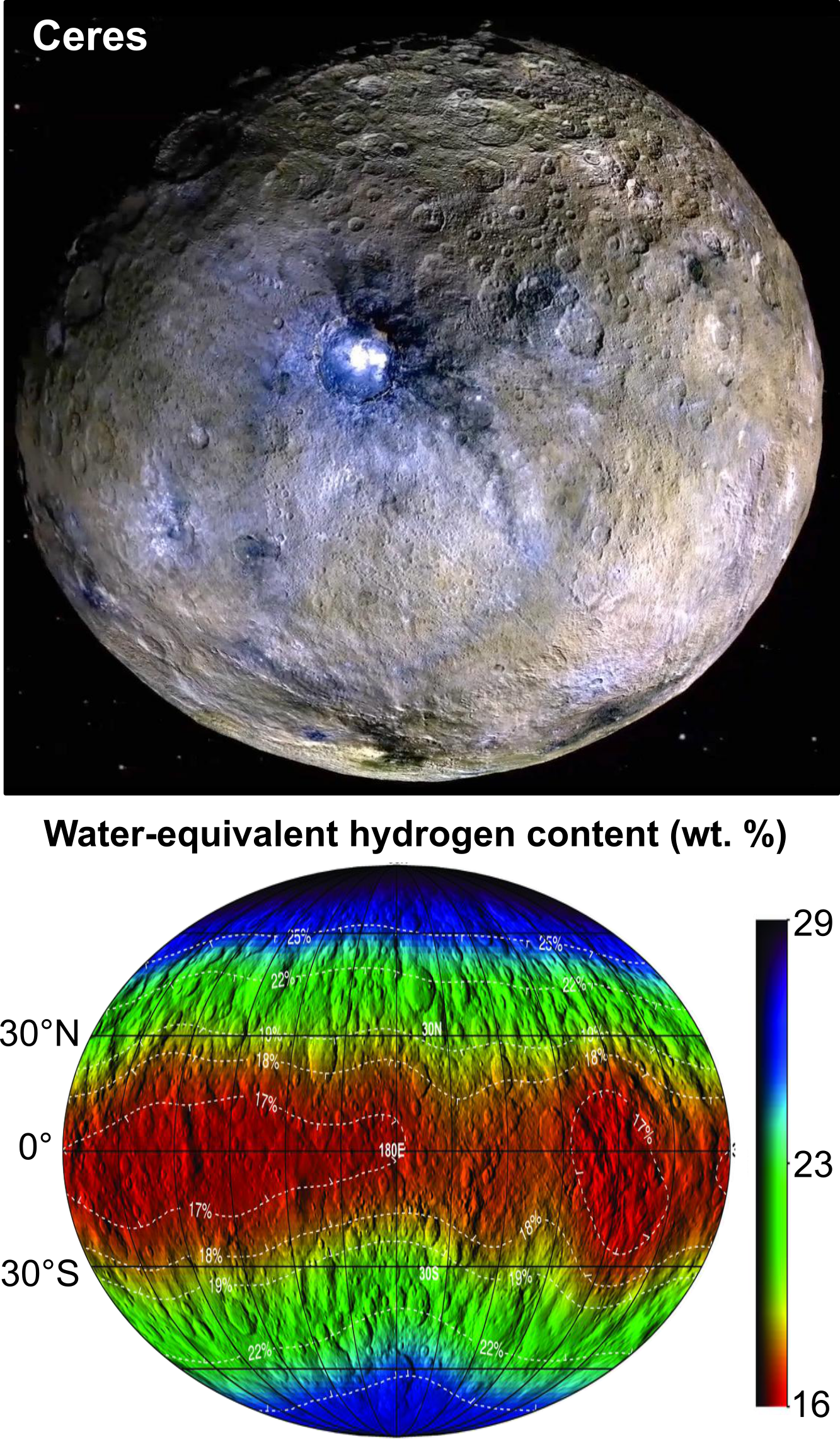}
			\caption{\textit{(top) False color Framing Camera mosaic of Ceres collected by NASA's Dawn spacecraft (credit: NASA, JPL-Caltech, UCLA, MPS, DLR, IDA, PIA20182). (bottom) Global map showing the distribution of H$_2$O-equivalent H in Ceres' regolith, derived from Gamma Ray and Neutron Detector (GRaND) data (modified from \citealt{prettyman2017CeresGRaND}). The mapped H is likely sourced from a subsurface H$_2$O ice layer that is exposed by impact events.}}
            \label{Ceres}
		\end{figure}
        
\textit{Diameter $\sim$939 km, angular diameter 0.34 -- 0.85''.} 

The Dawn spacecraft revealed that the largest main belt asteroid Ceres (Figure~\ref{Ceres}) is a possible ocean world \citep[e.g.,][]{castillo2020ceresOWs} that may have harbored a large volume of internal liquid H$_2$O in the past. Ceres likely retains a H$_2$O ice-rich layer protected beneath its regolith \citep[e.g.,][]{prettyman2017CeresGRaND}, which could contribute to a transient H$_2$O vapor exosphere, primarily due to impact exposure of this layer \citep{landis2019CeresH2Ovapor}. Although it is uncertain whether Ceres still has internal \textit{liquid} H$_2$O reservoirs, its surface exhibits recently-emplaced deposits of carbonate and chloride salts that are likely tracers of alteration in an aqueous environment \citep[e.g.,][]{desanctis2016CeresCO3, carrozzo2018CeresCO3, raponi2019Ceresmineralogy, bramble2022CeresNaCl}. Local exposure of hydrated NaCl, expected to dehydrate within years on Ceres’ surface, suggests ongoing surface eruption of subsurface brines \citep{desanctis2020CeresNaCl,nathues2020Cerescryo}. Organic-rich material detected on the surface of Ceres \citep[e.g.,][]{desanctis2017Ceresorganics} possibly originates from meteoritic infall and/or Ceres’ shallow interior \citep{bowling2020Ceresorganics}.
        
\subsubsection{Relevance to NASA Spacecraft Missions.} A New Frontiers-class Ceres lander and sample return mission concept \citep{castillo2022Cereslander} prioritized by the OWL Decadal Survey is a target on NASA’s New Frontiers 5 (NF5) list \citep{national2025NF5list}. If selected, the spacecraft would arrive in the mid to late 2030s or early 2040s, potentially overlapping HWO operations, which could provide complementary, high spectral resolution UV-NIR data of Ceres' transient exosphere. Should a Ceres mission be selected in a later competition, HWO could provide crucial continued observational coverage until this spacecraft arrives.
        
\subsubsection{Desired HWO Measurements.} 
\begin{itemize}
\item{Search for and measure sporadic H$_2$O vapor release, possibly due to impacts and strong sputtering events \citep{villarreal2017Ceresexosphere}.}

\item{Leverage the high SNR provided by HWO to search for and measure exospheric Na and other ions that originated in salts.}

        \begin{figure*}[h!]
            \begin{center}
			\includegraphics[scale=0.80]{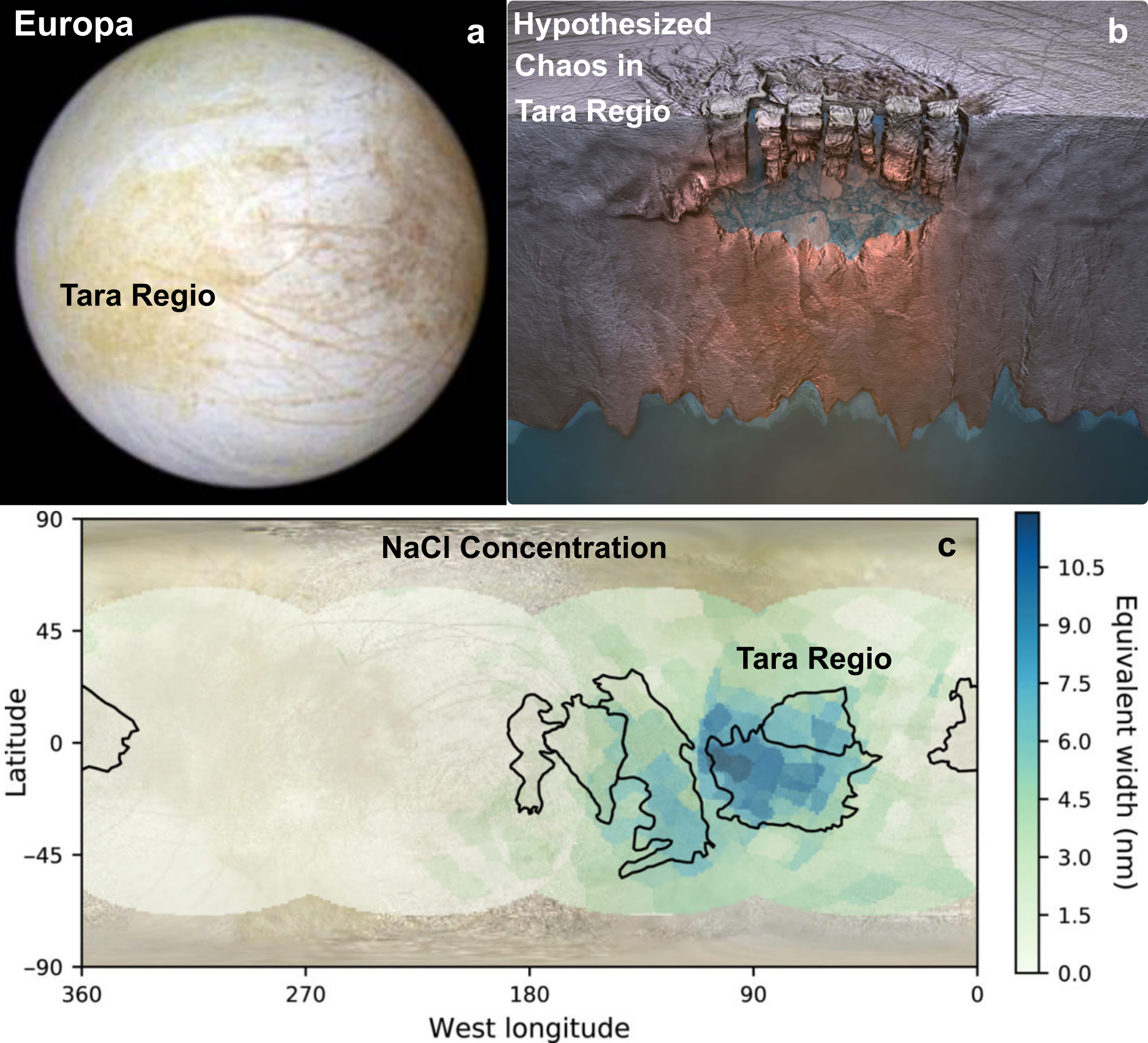}
			\caption{\textit{(a) Galileo Solid State Imaging (SSI) false color data of Europa's leading hemisphere and Tara Regio (credit: NASA, JPL/Caltech, PIA01295). (b) Illustration of chaos on Europa (credit: NASA, GSFC, PIA00840), showing hypothesized geologic communication between its subsurface ocean and surface (chaos formation mechanisms summarized in e.g., \citealt{daubar2024Clippergeol}). (c) Spectral map showing the distribution of irradiated NaCl on Europa, highlighting its concentration in and around Tara Regio (modified from \citealt{trumbo2019EuropaNaCl}).}}
            \label{Europa}
            \end{center}
		\end{figure*}

\item{Measure carbon-rich residues in the EUV and FUV \citep{hendrix2016ceresHST} and salt deposits at longer wavelengths.}
\end{itemize}

 
\subsection{Europa}
\textit{Diameter $\sim$3122 km, angular diameter 0.7 -- 1.05''.}       
        
Observations made by the Galileo spacecraft and HST indicate that Europa has a subsurface ocean that might interact with its surface (Figure~\ref{Europa}). Galileo imagery has revealed geologic landforms (e.g., domes, flow-like features, chaos terrains, etc.) that are indicative of communication between Europa's surface and liquid water in its interior \citep{greeley1998Europageol, greeley2000Europageol, fagents2003Europacryo, greeley2004Europageol, pappalardo1999Europaocean, kattenhorn2014Europasubsump}. In particular, Tara Regio on Europa’s leading hemisphere is dominated by chaos terrains and is apparently enriched in NaCl, H$_2$O$_2$, CO$_2$, and other components suggestive of recent activity and cycling of material between Europa’s subsurface, surface, and atmosphere \citep[e.g.,][]{trumbo2019EuropaNaCl, villanueva2023EuropaJWST, cartwright2025EuropaJWST}. Tara Regio and other chaos-dominated regiones may therefore represent ideal locations to assess the habitability of Europa's ocean beneath its surface \citep[e.g.,][]{hesse2022Europaexchange, chivers2023Europaexchange, vance2023Clipperhabitability}. HST/STIS data revealed H Lyman-alpha and oxygen O\textsuperscript{1} 130.4 nm emission, consistent with 200 km high H$_2$O vapor geyser emanating from Europa's south polar region \citep{roth2014Europaplume}. Subsequent studies have reported tentative supporting evidence for geyser activity and H$_2$O vapor \citep{sparks2016Europaplumes, sparks2017Europaplumes, jia2018Europaplume, paganini2020Europavapor}. However, more recent JWST observations found no trace of H$_2$O vapor, suggesting that geyser activity must be transient in nature \citep{villanueva2023EuropaJWST}.

\subsubsection{Relevance to NASA and ESA Spacecraft Missions.} NASA's Europa Clipper \citep{becker2024Clippercomposition, pappalardo2024Clippermission} and ESA's JUICE \citep{grasset2013JUICE, tosi2024JUICEcomp} spacecraft will arrive in the Jupiter system in the early 2030s. If the nominal missions are successful, extended missions could overlap HWO operations, providing for complementary, high R observations spanning the UV, VIS, and NIR.

        \begin{figure*}[!h]
            \begin{center}
			\includegraphics[scale=0.80]{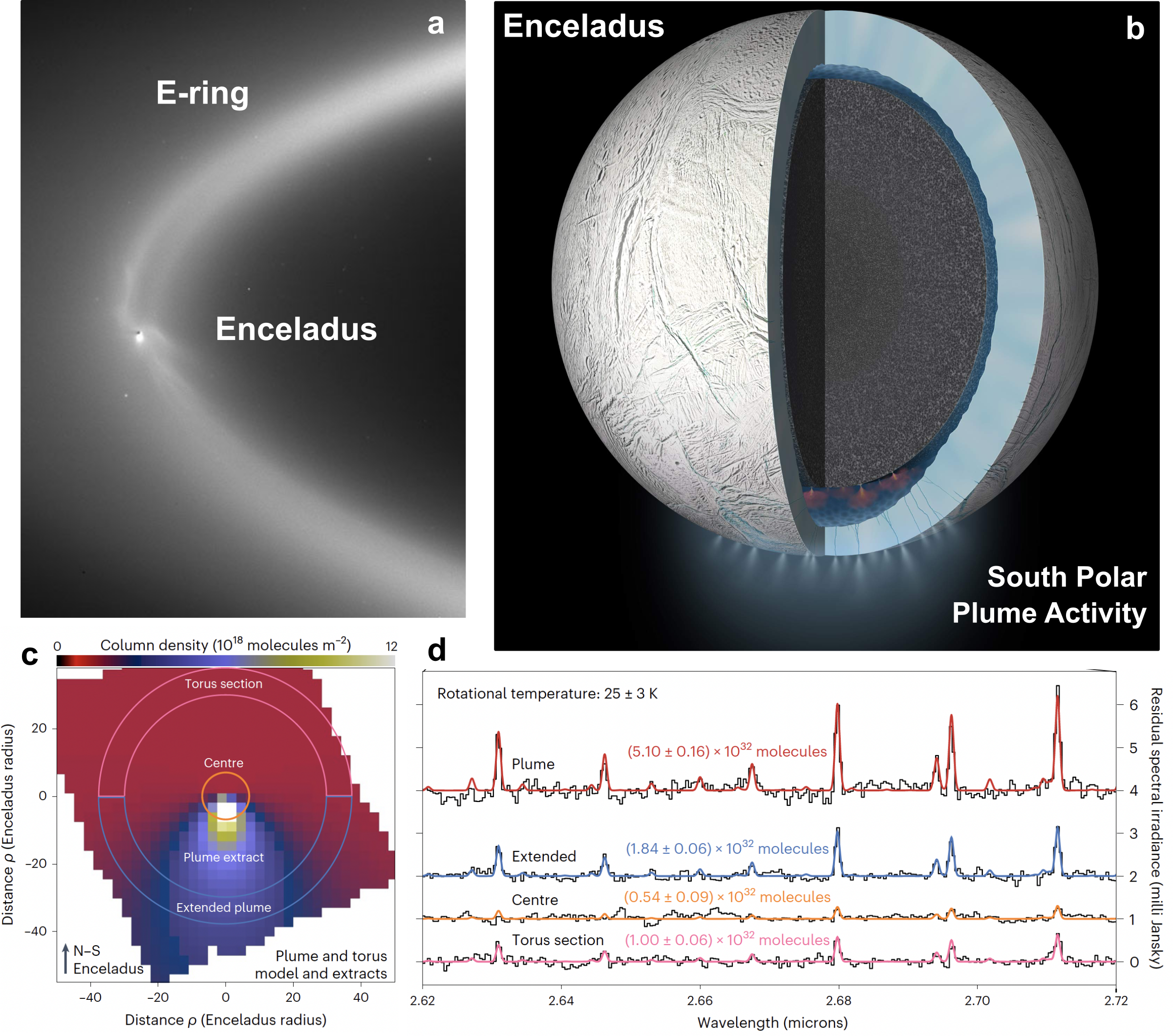}
			\caption{\textit{(a) Cassini Imaging Science Subsystem (ISS) image of Enceladus orbiting in the E-ring, which is sustained by icy material erupted from Enceladus' south polar geysers (credit: NASA, ESA, JPL, CICLOPS). (b) Cut away model showing Enceladus' interior, subsurface ocean, and geyser activity near its south pole (credit: NASA, JPL/Caltech, PIA20013). (c) JWST/NIRSpec IFU spectral cube highlighting the widespread signature of H$_2$O vapor around Enceladus \citep{villanueva2023EnceladusJWST}. (d) Extracted JWST/NIRSpec spectra exhibiting diagnostic H$_2$O vapor emission lines \citep{villanueva2023EnceladusJWST}.}}
            \label{Enceladus}
            \end{center}
		\end{figure*}
        
\subsubsection{Desired HWO Measurements.} 

\begin{itemize}
\item Confirm the presence of transient geysers and determine their eruptive cadence. HWO, as a serviceable facility that will be operational over multi-decadal timescales, would be ideal for capturing transient geyser activity at Europa.

\item Perform repeated observations at different time intervals to assess possible temporal changes in geologic activity and compare to spacecraft datasets (e.g., Galileo, Clipper, JUICE) and ground-based, HST, and JWST datasets. 

\item Assess the distribution of NaCl (e.g., UV and VIS color centers, NIR features) and other salts and explore compositional connections to Europa's exosphere (e.g., Na, K and other emission lines). 

\item Search for and measure carbon-rich surface residues in the EUV and FUV.

\end{itemize}
        
\subsection{Enceladus} 
        
\textit{Diameter \(\sim\)505 km, angular diameter 0.06 -- 0.09''.}

Observations made by Cassini detected active, large-scale geysers emanating from Enceladus’ south polar terrain (Figure~\ref{Enceladus}) \citep[]{porco2006EnceladusISS}. These geysers are primarily jetting H$_2$O vapor and ice grains into orbit, forming Saturn's E-ring. Other, more minor components include NH$_3$ \citep{waite2009EnceladusNH3}, macromolecular organics, phosphorus \citep{postberg2023EnceladusPO4}, and other key ingredients of habitability. Geyser activity on Enceladus is likely spurred by tidal heating of its interior that helps sustain a liquid water subsurface ocean. These geyser emissions vary with Enceladus' orbital position \citep{hedman2013Enceladusplume}, and long-term monitoring campaigns are needed to understand how internal tidal stresses help drive geyser activity and whether they change over multi-decadal timescales.
        
\subsubsection{Relevance to NASA and ESA Spacecraft Missions.} Enceladus geyser sampling mission concepts will be solicited as part of the NF5 call \citep{national2025NF5list}. An Enceladus south polar lander and an orbiter with geyser sampling capabilities has been identified as the next ESA Large class mission \citep{martins2024ESAVoyage2050}. Additionally, the Enceladus Orbilander \citep{mackenzie2021EncelOrbilander} is the second-priority Flagship mission concept for this decade, after a mission to the Uranian system \citep{national2022OWL}. These missions, if implemented, would arrive in the 2040s to early 2050s. HWO operations would possibly overlap with these missions, or provide crucial continued observational coverage until they arrive, thereby providing good opportunities for high R spectroscopy over UV-NIR wavelengths to complement spacecraft spectral measurements of Enceladus’ surface and its south polar gesyers.
        
\subsubsection{Desired HWO Measurements} 

\begin{itemize}
\item Long-term monitoring of geyser activity and comparison to archival datasets to determine if and how geyser intensity and composition change over time. HWO, as a serviceable facility that will be operational over multi-decadal timescales, would be ideal for monitoring geyser dynamics and evolution at Enceladus.

\item Search for and measure salts on Enceladus’ surface and ions in its torus / exosphere that are likely derived from salts, especially carbonates.

\item Better characterize the compositional evolution of material transported from Enceladus’ geysers to Saturn’s E-ring, which extends far beyond Enceladus’ surface, requiring an IFS.

\item Measure spectral tracers of other components originally detected by Cassini in Enceladus' geyser material and on its surface. 

\item Search for and measure carbon-rich surface residues in the EUV and FUV.

\end{itemize}
        
\subsection{The moons of Uranus, featuring Ariel}

\textit{Diameters 472 -- 1577 km, angular diameters 0.04 -- 0.12''.}
        
\begin{figure}[!b]
			\includegraphics[width=\columnwidth]{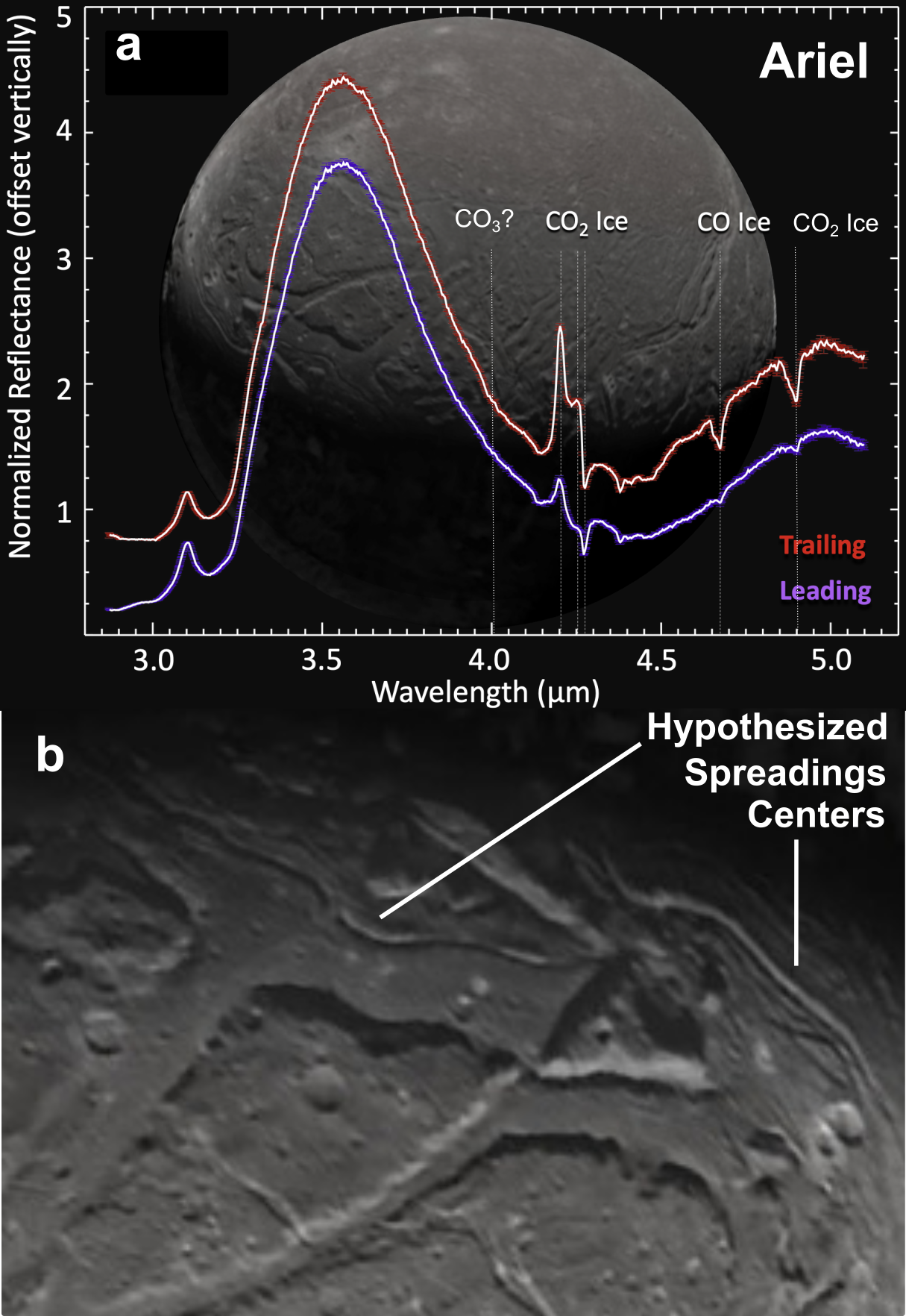}
			\caption{\textit{(a) JWST data showing spectral evidence for CO$_2$ and CO ices and possible CO$_3$-bearing salts that might have originated in Ariel's interior and reached its surface via geologic activity \citep{cartwright2024ArielJWST}. (b) Voyager 2 Imaging Science System (ISS) data showing possible spreading centers that may be conduits to Ariel's interior \citep{beddingfield2025Arielspreading}.}}
            \label{Ariel}
		\end{figure}

The large Uranian satellites are candidate ocean worlds \citep{cartwright2021Umoons} that may possess residual oceans in their interiors \citep{castillo2023oceansUmoons}. Whether these oceans can interact with the surfaces of these moons is uncertain, but intriguing evidence for past endogenic activity exists on these moons, especially on Ariel (Figure~\ref{Ariel}) where putative cryovolcanic features \citep{beddingfield2021Arielcryo} and interior-surface conduits \citep{beddingfield2025Arielspreading} have been identified. Furthermore, JWST has revealed the presence of CO ice, concentrated CO$_2$ deposits, and evidence for carbonate minerals on Ariel. These carbon oxides may in part originate in Ariel’s interior, possibly from a subsurface ``soda'' ocean that is rich in CO$_2$ and perhaps fairly acidic \citep{cartwright2024ArielJWST}. The hemispherical asymmetry in the distribution of the carbon oxides detected on Ariel (Figure~\ref{Ariel}) and the other large Uranian moons may result from radiolytic processing of native material derived from internal sources, primarily on their trailing hemispheres \citep[e.g.,][]{grundy2006UmoonsCO2, cartwright2015UmoonsCO2}.

\subsubsection{Relevance to NASA Spacecraft Missions} The Uranus Orbiter and Probe (UOP) mission concept \citep{national2022OWL} would arrive in the Uranus system after HWO is operational (\(\sim\)2050), providing important opportunities for complementary high R, UV-NIR spectroscopy of Uranus’ moons, rings, and the planet itself. For example, HWO could characterize the surface compositions of the moons in the 2040s, before their north polar regions are obscured by winter darkness (subsolar point migrates from 82°S to 82°N over the course of Uranus' 84-year orbit). These observations would have important implications for understanding seasonality at Uranus and the associated migration of CO$_2$, CO, and other volatiles across the moons’ surfaces in response to changes in subsolar heating. Furthermore these HWO data would be collected in the runup to equinox in 2050, capturing a key temporal snapshot that UOP will likely miss.
        
\subsubsection{Desired HWO Measurements.} 

\begin{itemize}
\item Search for and establish long-term monitoring of Ariel’s predicted exosphere that forms from seasonal sublimation \citep{grundy2006UmoonsCO2, sori2017UmoonVoltrans, steckloff2022voltrans, menten2024voltrans} and sputtering \cite[e.g.,][]{raut2013CO2sputtering} of CO, CO$_2$, and other surface volatiles with an IFS.

\item Determine whether CO$_2$ and other carbon oxides detected on Ariel originate from its interior or are primarily formed via irradiation processing of its surface. 

\item Search for and measure carbon-rich surface residues in the EUV and FUV. 

\item Search for and measure geyser activity at Ariel.

\end{itemize}

        \begin{figure}[!b]
			\includegraphics[width=\columnwidth]{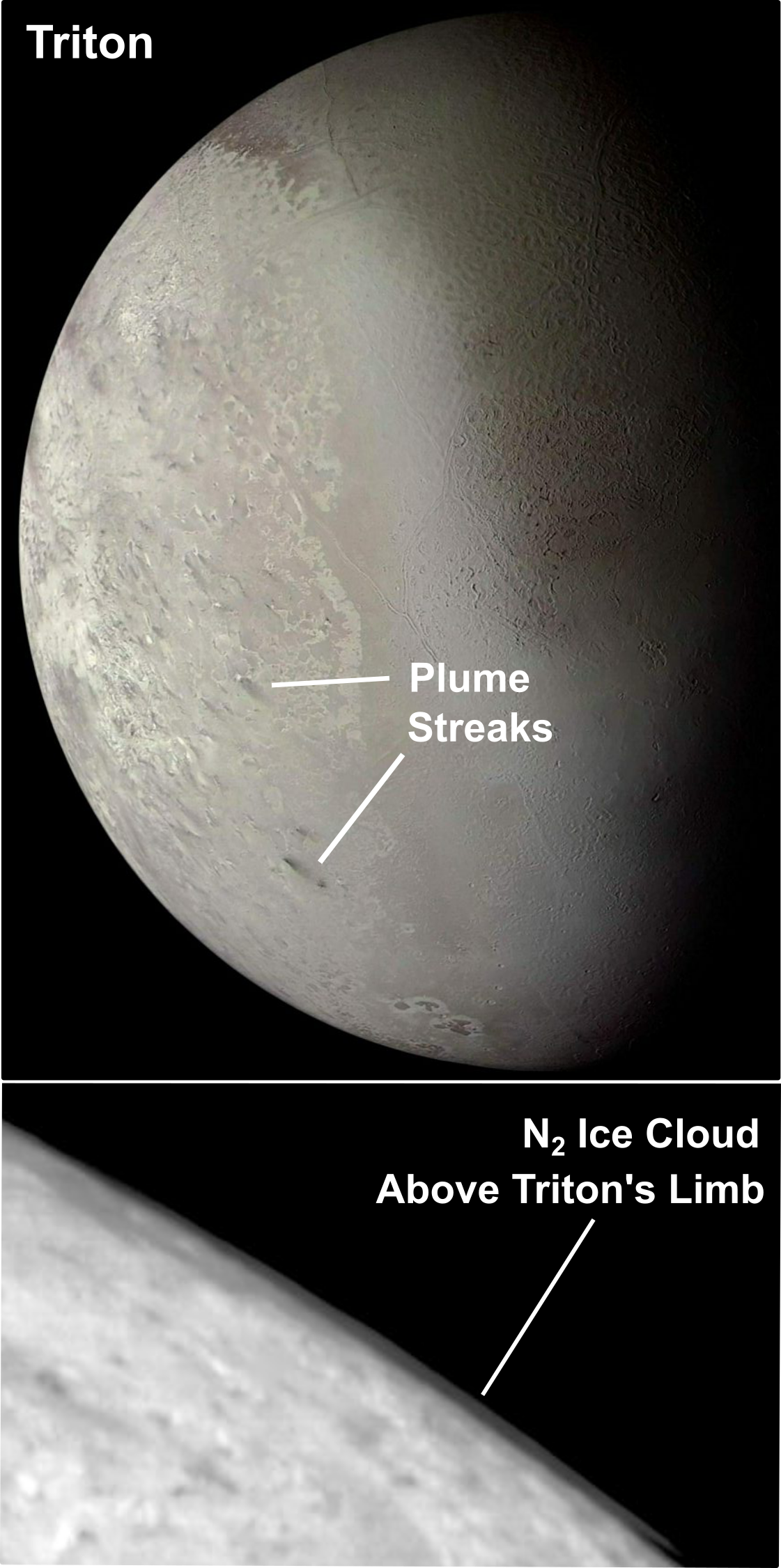}
			\caption{\textit{(top) Voyager 2 ISS image mosaic showing the variety of geologic regions and geyser streaks on Triton. (bottom) Close-up showing a thin N$_2$ ice cloud imaged by Voyager 2 \citep{smith1989NeptuneV2}.}}
            \label{Triton}
		\end{figure}

\subsection{Triton}

\textit{Diameter $\sim$2706 km, angular diameter 0.12 -- 0.13''.}

Neptune's moon Triton (Figure~\ref{Triton}) is very likely a captured trans-Neptunian object (TNO) \citep{agnor2006Tritoncapture} with a surface rich in volatile ices \citep[e.g.,][]{cruikshank1993Tritonices} that migrate across Triton over seasonal and longer timescales \citep[e.g.,][]{bertrand2022Tritonvolatiles}. Voyager 2 imaged active geysers and streaks from past geyser activity \citep{smith1989NeptuneV2}, possibly driven by endogenic geologic activity, or alternatively, sublimation processes \citep[e.g.,][]{hofgartner2022Tritonplume}. Voyager 2 also imaged thin clouds in a tenuous N$_2$ atmosphere \citep[e.g.,][]{broadfoot1989TritonUV} over Triton's limb \citep{smith1989NeptuneV2}, highlighting the high level of activity on this very cold icy body (estimated mean surface temperatures $<$ 40 K; \citealt{conrath1989NeptuneV2}). 

The evidence of recent and ongoing activity and volatile ices that may be replenished from Triton's interior make it a candidate ocean world \citep[e.g.,][]{hansen2021TritonOW}. Beyond Triton, several large TNOs such as Pluto, Eris, and Makemake may be able to hold onto enough radiogenic heat to maintain liquid environments in their interiors \citep[e.g.,][]{kamata2019Plutoocean}.

\subsubsection{Relevance to NASA Spacecraft Missions.} A Triton Ocean World Surveyor concept was prioritized for a late-2020s / early-2030s NF Announcement of Opportunity \citep{national2022OWL}, and a Discovery-class Triton flyby mission was recently proposed \citep{prockter2022Trident}. HWO would enable continuous monitoring of Triton before a dedicated mission arrives. 

Dedicated missions to large TNOs are not planned at this time. The wealth of discoveries from the New Horizons mission to Pluto \citep{stern2018plutoNH}, as well as discoveries made with JWST \citep[e.g.,][]{glein2024ErisMakemake, kiss2024MakemakeJWST}, suggest trans-Neptunian dwarf planets are well worth monitoring for compositional characterization and activity on multi-decadal timescales. 
        
\subsubsection{Desired HWO Measurements (includes Pluto, Eris, Makemake, and other large TNOs)} 

\begin{itemize}
\item Establish long-term monitoring of Triton's surface and atmospheric composition. 

\item Determine whether internally-derived species are present on Triton's surface.

\item Search for large-scale compositional changes on the surface of Triton over multi-decadal timescales.
\end{itemize}

\begin{table*}[!h]
\caption{\textbf{Physical parameters of ocean worlds required to address the scientific objectives.}}
\label{Table 1}
{\vspace{0.1cm}\small
\begin{tabular}{lllll}  
\tableline
\noalign{\smallskip}
	\begin{tabular}[c]{@{}l@{}}Physical \\ Parameter \end{tabular} & 
	\begin{tabular}[c]{@{}l@{}}State-of-the-Art \\ from Existing Data\end{tabular} &
	\begin{tabular}[c]{@{}l@{}}Incremental Progress \\(Enhancing) \end{tabular} &     \begin{tabular}[c]{@{}l@{}}Substantial Progress \\(Enabling) \end{tabular} &  
	\begin{tabular}[c]{@{}l@{}}Major Progress \\(Breakthrough) \end{tabular} \\
\noalign{\smallskip}
\tableline
\noalign{\smallskip}
\begin{tabular}[c]{@{}l@{}}Geyser activity \\ and composition: \\Searches, \\measurements, \\ and monitoring \end{tabular} & \begin{tabular}[c]{@{}l@{}}JWST and Cassini \\ measured H$_2$O\\ vapor at Enceladus; \\ HST detected O \\ emission at Europa;\\ Voyager 2 imaged\\ geysers at Triton \end{tabular} & \begin{tabular}[c]{@{}l@{}}Determine if Enceladus\\ geysers change over \\ time; confirm Europa \\ geysers; measure the \\ composition of \\ Triton's geysers \end{tabular} & \begin{tabular}[c]{@{}l@{}}Measure non-H$_2$O gases \\in Enceladus geysers \\(CO$_2$, NH$_3$, H$_2$, etc); \\Determine Europa's\\ geyser composition and \\ eruption cadence; detect \\ geysers elsewhere \end{tabular} & \begin{tabular}[c]{@{}l@{}}Define the spectrum\\ and sources of geyser\\ activity across the \\Solar System (spatial \\distribution, temporal \\variation, composition)\\ \end{tabular} \\
\noalign{\smallskip} 
\tableline
\noalign{\smallskip} 
\begin{tabular}[c]{@{}l@{}}Surface \\ composition: \\ Identify and \\ measure \\ocean-derived \\ components \end{tabular}& \begin{tabular}[c]{@{}l@{}}Distribution of NaCl \\and association with \\young geologic terrains \\on Europa; detection\\ of NH$_4$ and carbonates\\ on Ceres \end{tabular}& \begin{tabular}[c]{@{}l@{}}Detection and \\measurement of \\salts and other\\ ocean-derived\\ products on \\other icy bodies \end{tabular}& \begin{tabular}[c]{@{}l@{}}Determine which\\ specific salts (carbonates,\\ sulfates, chlorides, etc)\\ are present and use to \\infer ocean conditions\\ and geochemistry (pH, etc)\end{tabular}& \begin{tabular}[c]{@{}l@{}}Confirm linkage between\\ astrobiologically-relevant\\ components (organics, \\ phosphates, etc) and\\ geologic conduits to\\ subsurface oceans \end{tabular}\\
\noalign{\smallskip} 
\tableline
\noalign{\smallskip} 
\begin{tabular}[c]{@{}l@{}}Exosphere \\ composition: \\ Identify, measure \\ ocean-derived \\ components \end{tabular} & \begin{tabular}[c]{@{}l@{}}Detection of Na and K \\ in Europa’s exosphere,\\ H$_2$O vapor at \\ Enceladus; CO$_2$ gas \\ at Ganymede and \\ Callisto; N$_2$ at Triton \end{tabular} & \begin{tabular}[c]{@{}l@{}}Detection of ions, \\ and molecular \\ gases at the Uranian \\ moons and other \\icy bodies \end{tabular} & \begin{tabular}[c]{@{}l@{}}Long-term monitoring \\of exospheres and \\mapping of gas \\distributions to \\investigate volatile \\cycling \end{tabular} & \begin{tabular}[c]{@{}l@{}}Ascertain composition, \\ sources, and sinks of \\ exospheric species at \\ ocean worlds across \\ the Solar System \end{tabular} \\
\noalign{\smallskip}
\tableline\
\end{tabular}
}
\end{table*}

\begin{table*}[!h]
\caption{\textbf{Future HWO observations required to address the science objectives.}}
\label{Table 2}
\begin{center}
{\vspace{-0.6cm}\small
\begin{tabular}{lllll}  
\tableline
\noalign{\smallskip}
	\begin{tabular}[c]{@{}l@{}}Observation \\ Requirement \end{tabular} & 
	\begin{tabular}[c]{@{}l@{}}State of the Art \end{tabular} &
	\begin{tabular}[c]{@{}l@{}}Incremental Progress \\(Enhancing) \end{tabular} &     \begin{tabular}[c]{@{}l@{}}Substantial Progress \\(Enhancing) \end{tabular} &  
	\begin{tabular}[c]{@{}l@{}}Major Progress \\(Breakthrough) \end{tabular} \\
\noalign{\smallskip}
\tableline
\noalign{\smallskip}
\begin{tabular}[c]{@{}l@{}}UV spectroscopy \end{tabular} & \begin{tabular}[c]{@{}l@{}}HST, spacecraft \\ (e.g. Clipper UVS) \end{tabular} & \begin{tabular}[c]{@{}l@{}}90 -- 200 nm; \\ R $\sim$1000;\\ SNR $>$ 100 for\\ Vmag 10 icy body \end{tabular} & \begin{tabular}[c]{@{}l@{}}90 -- 300 nm; \\ R $\sim$3000;\\ SNR $>$ 100 \\ for Vmag 15 icy body \end{tabular} & \begin{tabular}[c]{@{}l@{}}50 -- 400 nm (capture irradiated \\ carbon features in EUV); \\ R $\sim$10,000 (Resolve emission lines\\ for Mg and other ions);\\ SNR $>$ 100 for Vmag 20 icy body \end{tabular} \\
\noalign{\smallskip} 
\tableline
\begin{tabular}[c]{@{}l@{}}VIS spectroscopy \end{tabular} & \begin{tabular}[c]{@{}l@{}}HST, spacecraft \\ (e.g. Cassini VIMS) \end{tabular} & \begin{tabular}[c]{@{}l@{}}400 -- 700 nm; \\ R $\sim$1000;\\ SNR $>$ 100 \\ for Vmag 15 icy body; \\ large dynamic range \\ to capture some \\ brighter targets\end{tabular} & \begin{tabular}[c]{@{}l@{}}350 -- 1100 nm; \\ R $\sim$3000;\\ SNR $>$ 100 \\for Vmag 20 icy body; \\ large dynamic range, \\saturation-mitigation \\ to partially capture \\ all bright targets \end{tabular} & \begin{tabular}[c]{@{}l@{}}350 -- 1100 nm; \\ R $\sim$10,000 (fully resolve emission \\ lines for Na and other ions);\\ SNR $>$ 100 for Vmag 20 icy body \\ large dynamic range and \\ saturation-mitigation techniques to \\fully capture all bright targets\end{tabular} \\
\noalign{\smallskip} 
\tableline
\begin{tabular}[c]{@{}l@{}}NIR spectroscopy \end{tabular} & \begin{tabular}[c]{@{}l@{}}JWST, spacecraft \\ (e.g. Clipper MISE) \end{tabular} & \begin{tabular}[c]{@{}l@{}}800 -- 2500 nm; \\ R $\sim$1000; (detect broad \\ salt and ice features);\\ large dynamic range to \\ capture some bright \\ targets; cryocooled \\ detector ($\sim$90 K) \end{tabular} & \begin{tabular}[c]{@{}l@{}}800 -- 3000 nm; \\ R $\sim$3000;\\ large dynamic range,\\ saturation-mitigation \\ to partially capture \\all bright targets;\\ cryocooled detector\\ ($\sim$60 K) \end{tabular} & \begin{tabular}[c]{@{}l@{}}800 -- 5000 nm; \\ R $\sim$10,000; (Fully resolve H$_2$O \\ vapor and other gas features);\\ large dynamic range and \\ saturation-mitigation techniques \\ to fully capture all bright targets;\\ cryocooled detector \\ ($\sim$30 K) \end{tabular} \\
\noalign{\smallskip} 
\tableline
\begin{tabular}[c]{@{}l@{}}IFS and imager \\ angular resolution \end{tabular} & \begin{tabular}[c]{@{}l@{}}Numerous \\ spacecraft \end{tabular} & \begin{tabular}[c]{@{}l@{}}$\sim$0.02''/pixel \\ spatial scale; \\large dynamic range \\ to capture some \\ bright targets \end{tabular} & \begin{tabular}[c]{@{}l@{}}$\sim$0.015''/pixel \\ spatial scale; \\large dynamic range, \\saturation-mitigation \\ to partially capture all \\ bright targets \end{tabular} & \begin{tabular}[c]{@{}l@{}}$\sim$0.01''/pixel \\ spatial scale; \\large dynamic range, \\saturation-mitigation techniques \\to fully capture all bright targets \end{tabular} \\
\noalign{\smallskip} 
\tableline
\end{tabular}
}
\end{center}
\end{table*}

\section{Physical Parameters and Description of Observations}

\textit{Primary Objectives.} Use an IFS to observe a large suite of ocean worlds and monitor long-term temporal changes in their surface and exospheric compositions, the cadence of geyser activity, and the column densities of detected gases. These observations would ideally span a wide wavelength range (50 -- 5000 nm) with high sensitivity (SNR >100) and high spectral resolution (R $\sim$10,000). The physical parameters and desired observations required to achieve these objectives are summarized in Tables 1 and 2, respectively. Below and in Table 1, we provide some hypothetical examples of incremental, substantial, and major progression in our understanding of ocean world habitability that could be achieved with an IFS on HWO.

\textit{State of the Art (SOA):} Multiple Cassini instruments determined that Enceladus is geologically active and has a subsurface ocean source for its observed geysers. HST observations suggest that Europa exhibits south polar geyser activity as well \citep{roth2014Europaplume}. Similarly, Voyager 2 revealed that Triton was recently geologically active, including visual evidence for past geyser activity \cite{smith1989NeptuneV2}. However, confirming geyser activity at Europa with follow up observations has been challenging, and whether Triton's geysers are driven by endogenic heating similar to Enceladus is uncertain, as is the extent of interactions between Triton's proposed subsurface ocean and its surface \citep{hofgartner2022Tritonplume}.

\textit{Enhancing (incremental progress):} Long-term monitoring of Europa (UV-NIR) is needed to investigate the extent of surface-interior communication, primarily by determining how the spectral properties of its surface and exosphere changes over time. Confirmation of putative geyser activity at Europa is also required to investigate interior-surface communication. The collected compositional information could be used to inform geochemical models of Europa's interior and improve constraints on its ocean pH and other properties. Similar long-term monitoring of Ceres is required to search for recently exposed liquid and solid-state H$_2$O and hydrated salts on its surface, as well as H$_2$O vapor in its transient exosphere. Long-term monitoring of Enceladus’ geyser activity is needed to determine whether the total erupted flux and ratio between H$_2$O and non-H$_2$O components change over time, in particular for components relevant to deciphering the habitability of its internal ocean.

\textit{Enabling (substantial progress):} Long-term monitoring is required to enable detection and characterization of multiple geyser outbursts at Europa, which are likely sporadic and transient in nature. To this end, UV observations will be crucial for measuring gaseous and ionized components derived from Europa's geysers, and VIS and NIR observations will be key for measuring icy particles. NIR observations may also reveal regional-scale hot spots ($\gtrsim$500 km diameter) associated with geyser activity. Additionally, observing volatile cycling on Triton (UV-NIR) is required to place improved constraints on whether geyser activity is endogenic or solar-driven. Monitoring seasonal changes in the Uranian moons’ exospheres (UV-NIR) as their winter poles are exposed to sunlight during equinox is required to understand volatile transport on these moons.

\textit{Breakthrough (major progress):} Long-term monitoring is required to fully determine the frequency, location, and composition of geyser eruptions on Europa and determine whether geyser material originates from its subsurface ocean or from warm H$_2$O melt pockets in its icy crust. Long-term monitoring is also required to determine if the composition of Europa's geysers change over time or are compositionally distinct in different regions. Additionally, detection and characterization of geologic activity associated with internal oceans at large TNOs such as Pluto, Eris, and Makemake would be critical for defining the full suite of ocean worlds across the Solar Syste,, with important implications for understanding habitability on icy exoplanets far beyond their stars' H$_2$O snowlines.
        
\subsection{Post-observation Analyses}

\begin{itemize}
\item Compare HWO spectra to laboratory data and spectral models to determine which compounds are present in the exospheres and on the surfaces of these bodies, constrain rates of emplacement and loss, and assess relevance to habitability.

\item Compare VIS images to UV and NIR spectral cubes to map the distribution of constituents and assess association with specific geologic features and terrains on spatially resolved targets, such as the Galilean satellites.

\item Compare ocean worlds across the Solar System to assess the range of exhibited activity and improve understanding of the conditions required for habitable subsurface environments.
\end{itemize}

        \begin{figure*}[!h]
            \begin{center}
			\includegraphics[scale=0.50]{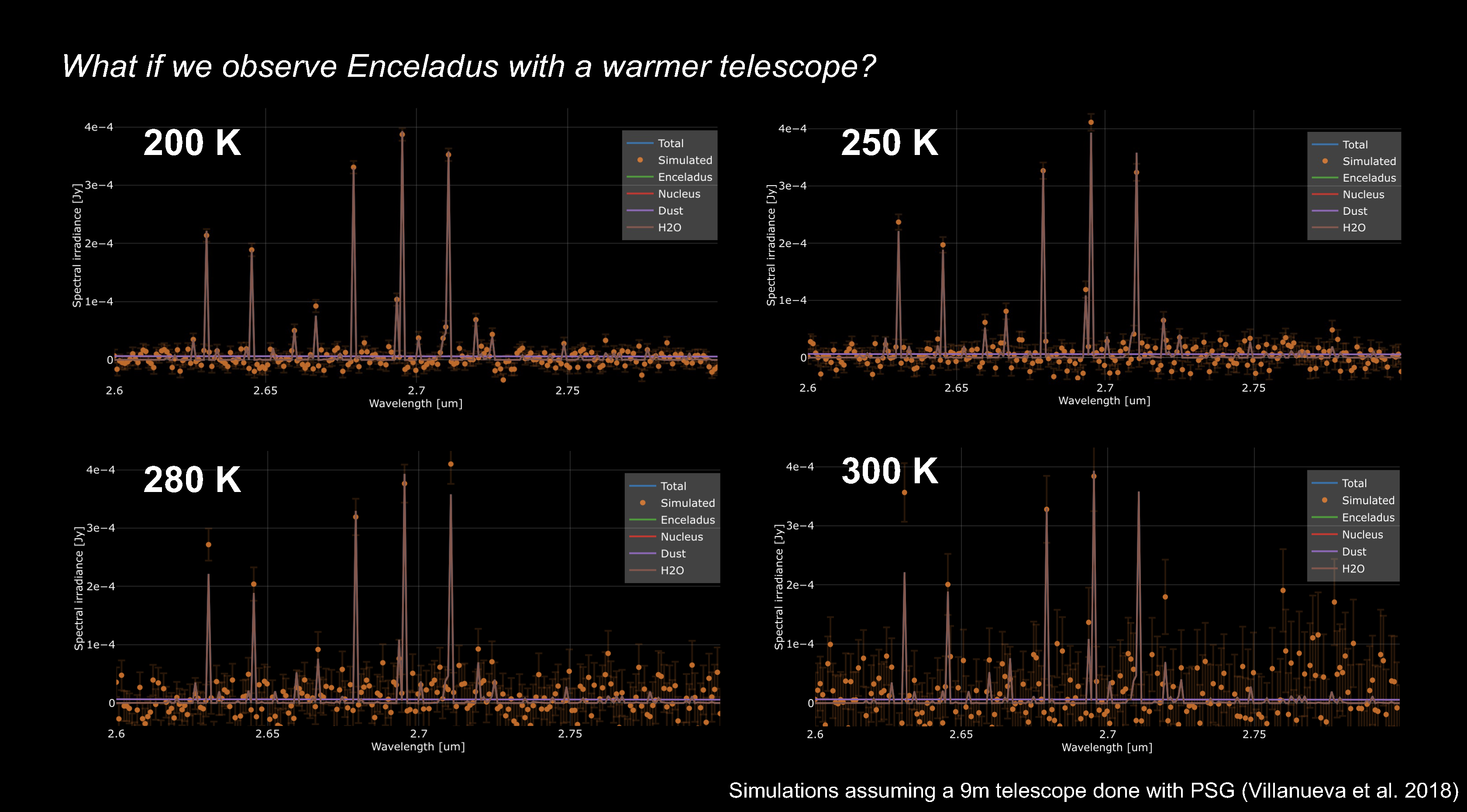}
			\caption{\textit{Simulations of H$_2$O vapor emission lines between 2600 and 2700 nm at Enceladus as a function of temperature, using a space telescope with a 9 m aperture. All plots were made using the Planetary Spectrum Generator \citep{villanueva2018PSG}.}}
            \label{Enceladus_warm}
            \end{center}
		\end{figure*}
        
\subsection{NIR spectroscopy with a warm telescope}
Ground-based NIR spectral observations conducted at room temperature with cryocooled detectors have provided important results for a wide range of Solar System targets. JWST is highlighting how important NIR observations are for understanding ocean world chemistry, in particular in the strong telluric bands between 2550 and 2850 nm and 4200 and 4500 nm where Earth’s atmosphere is opaque and rovibrational lines for H$_2$O vapor and CO$_2$ gas cannot be studied.

For example, erupted H$_2$O vapor from Enceladus’ geysers can be well-studied using space-based telescopes, as demonstrated by JWST. A NIRSpec-like spectrograph on a space telescope operating at warmer temperatures can still detect and measure H$_2$O vapor at Enceladus (Figure~\ref{Enceladus_warm}).


\acknowledgements
This publication was supported by investments of internal development funds from the Johns Hopkins University Applied Physics Laboratory. Part of this work was developed at the Jet Propulsion Laboratory, California Institute of Technology, under contract to NASA.


\newpage
\bibliography{Cartwright_HWO_OW-habitability}

\end{document}